\documentclass[aps,prb,twocolumn,groupedaddress,showpacs]{revtex4}
\usepackage{graphicx}
\usepackage{dcolumn}
\usepackage{bm}

\begin{document}

\title{Study of phase stability in the $\sigma$-FeCr system}

\author{J. Cieslak}
\email[Corresponding author: ]{cieslak@novell.ftj.agh.edu.pl}
\author{J. Tobola}
\author{S.M. Dubiel}
\affiliation{Faculty of Physics and Applied Computer Science,
AGH University of Science and Technology, al. Mickiewicza 30, 30-059 Krakow, Poland}

\date{\today}

\begin{abstract}
Formation energy of the $\sigma$-phase in the Fe-Cr alloy system, $\Delta E$,
was computed versus the occupancy changes on each of the five possible lattice sites.
Its dependence on a
number of Fe-atoms per unit cell, $N_{Fe}$, was either monotonically increasing
or decreasing function of $N_{Fe}$, depending on the site on which Fe-occupancy
was changed. Based on the calculated $\Delta E$ – values, the average formation
energy, $<\Delta E>$, was determined as a weighted over probabilities of
different atomic configurations. The latter has a minimum in the concentration
range where the $\sigma$-phase exists. The minimum in that range of
composition was also found for the free energy calculated for 2000 K
and taking only the configurational entropy into account.
\end{abstract}

\pacs{
      71.15.Mb,
      71.15.Nc,
      71.23.-k,
      75.50.Bb
      }

\maketitle

Sigma ($\sigma$) phase was found in 43 binary alloys and many
other three- or multi-component alloy systems \cite{Hall66,Joubert08}.
This phase cannot be formed at the stage of solidification of the solution
of alloying elements. Instead, it can only be obtained by a high temperature annealing
process (solid state reaction). From the viewpoint of technological applications of alloys, the $\sigma$-phase is
the one which should be avoided as it drastically deteriorates different mechanical properties of materials
in which it precipitates.
Sigma phase has a tetragonal unit cell (type D$^{14}_{4h}$ P$_4$2/mnm) hosting 30 atoms
distributed over five crystallographically non-equivalent sites usually called A, B, C, D, and E.
In binary alloy systems, like Fe-Cr - the subject of the present paper- both alloying elements are present on all five lattice sites \cite{Cieslak08c}.

In the Fe-Cr system the $\sigma$-phase can be formed by an isotermal annealing in a limited range of concentration
($\sim$45-50at\% Cr) and temperature ($\sim 500-830^\circ$C)\cite{Hall66}.
Once formed it remains stable at lower temperatures, but it
dissolves into the $\alpha$-phase at temperatures above $830^\circ$C. Details of the reasons and
mechanism of its creation as well as that of its dissolution, hence
prevention of its formation, are not known with sufficient clarity yet.

Determination of the formation energy of this phase compared to the energy of
formation of {\it bcc} ($\alpha$) phase (from which it precipitates) may
allow for a better understanding of processes involved in its formation and dissolution. Thus, it may be helpful
in a creation of a new generation of materials, like stainless steels, having properties more suitable for a construction
of new generations of industrailly important facilities like nuclear power plants.

Several theoretical papers were recently devoted to the $\alpha-\sigma$ phase transformation \cite{Turchi94,Olsson03,Korzhavyi05,Ackland09,Kabliman09},
yet the mechanism responsible for the process
is not completely understood. This justifies further studies toward this end.
The increase of computing capabilities of modern computers
allows creation of increasingly sophisticated models that are able to include more parameters relevant to more adequately and precisely describe
the complex structure and properties of the $\sigma$-phase.

In the present work we analysed semi-ordered unit cells having a composition in the vicinity of
experimentally determined concentration of one of the alloying elements.
Assumptions of the model and details
of a computational method used in this work have been widely described elsewere\cite{Cieslak08b,Cieslak10a}.

Determination of the formation energy of the $\sigma$-phase has so far been based on
calculations carried out for the unit cell which sublattices were occupied
entirely by a single type of atoms \cite{Sluiter75,Pavlu10}.  In the case of five sublattices and two
kinds of atoms, there are 32 possible different atomic arrangements which
satisfy these assumptions.  Differences between formation energies (in fact, free entalpies, $H$)
determined
for each of these configurations and corresponding values obtained for
pure constituents (so called reference state - RS),
$\Delta E = H^{\sigma}-H^{RS}$,
are customarily presented in form of a diagram as a function of total concentration of
alloying elements.  Such presentation enables a determination of the concentration range
where the $\sigma$-phase exists, as well as drawing conclusions regarding sublattice
occupancies.  Although this approach allows performing calculations without losing the
symmetry of the unit cell, it does not permit to take into account a sublattice disorder.
For example in Ref. 12, the lowest value of the formation energy was found for a system
named by the authors as FeCrCrFeCr.  This notation means that sublattices A and D are fully occupied
by Fe atoms, whereas B, C and E sites exclusively by Cr atoms.  In the present paper, the convention of
the description of the sublattice occupancy in the unit cell is slightly
different, because we want to take into account atomic chemical disorder, too.
Since the total number of atoms on each of the five sublattices is known, in our notation
only the number of the Fe atoms on each sublattice is given explicitely, while the Cr atoms
form the balance.  Thus, according to our convention the arrangement of
atoms FeCrCrFeCr will be called $\sigma$-20080.

Analysis of the formation energy, as sought here, is fraught with a lot of problems.
The most important seems to be the one that the $\sigma$-phase in the Fe-Cr
alloy system is chemically disordered, and all sublattices are occupied by both alloying
elements \cite{Cieslak08c}.
As found with neutron diffraction studies, sublattices A
and D are actually in the majority filled by Fe atoms while the other sublattices
in the majority by Cr atoms. Yet, for the Fe-Cr the average sublattice occupancy is
closer to the configuration $\sigma$-21373 rather than $\sigma$-20080.

Another serious problem that we encounter analyzing the $\Delta E's$ is related with the
fact that such analysis is performed as a function of the total
concentration of one of the alloying elements.  In fact, the $\Delta E's$ should be
better considered in five dimensions i.e.versus the concentration for each of the
five sublattices.

Facing these problems we proposed a different approach which so far has
been successfully applied to determine different hyperfine parameters of the $\sigma$-phase in
Fe-Cr and Fe-V systems, such as charge density, electric field gradients and magnetic
structure of particular sublattices.  It is based on the calculations
performed in a finite number of unit cells with the symmetry reduced to a simple tetragonal one,
in which each position (but not the sublattice) is occupied by one type of the atoms (e.g. Fe or Cr).
The atoms are distributed over
different sites with probabilities determined from neutron diffraction
measurements.  Details of the method itself and the mentioned-above calculated hyperfine
parameters are presented elsewhere \cite{Cieslak08b,Cieslak10a}.

\begin{figure}[bt]
\includegraphics[width=.40\textwidth]{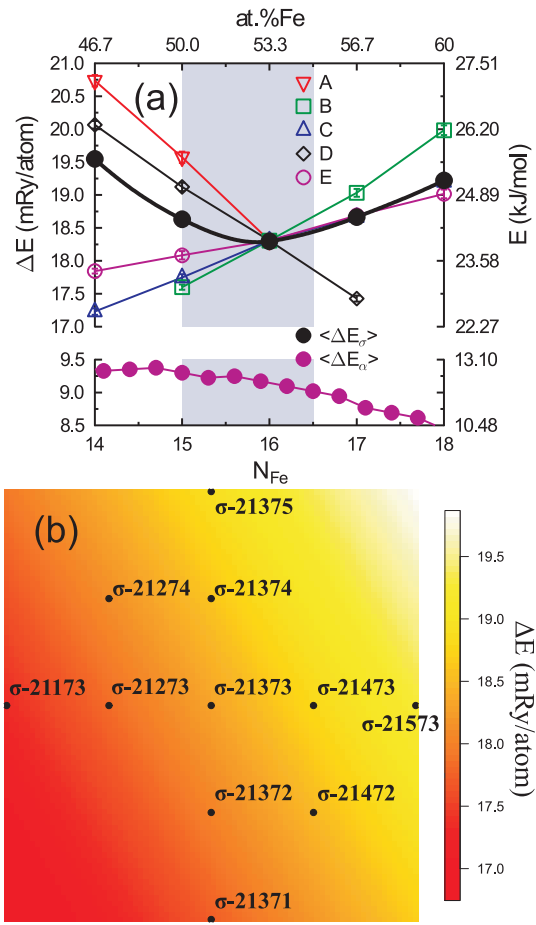}
\caption{(Online color)
(a) A change of the formation energy, $\Delta E$, per one atom versus a number of Fe-atoms per unit cell, $N_{Fe}$, for
different lattice sites.
$<\Delta E>$-values for both phases are shown, too. Solid lines are to guide the eye only. The blue band
markes a concentration range of the experimentally found range of the $\sigma$-phase occurance. A different representation of the C- and E-sublattice  data
displayed for the $\sigma$-phas in (a) is given in (b), where, additionally, an effect of a simultanous exchange of 2 atoms on different sublattices
(namely, $\sigma$-21274 and $\sigma$-21472) is visualized.
}
\label{fig1}
\end{figure}

Let assume the $\sigma$-21373 be a reference state for further calculations.  This
configuration was chosen as equivalent to the average Fe-occupancy of the sublattices
determined experimentally.  The value of $\Delta E$ for the $\sigma$-21373 was here calculated as the
average over the values computed for 26 unit cells of different configurations of
elementary atoms on all sublattices that fulfil the assumptions for the
$\sigma$-21373.  The next step was to calculate a change of $\Delta E$ caused by a
replacement of one or two atoms of Fe (Cr) but on a single sublattice, only.

The calculated in that way values
of $\Delta E$ are shown in FIG 1a. Each point on the chart corresponds to 10-15 various
atomic arangements analysed separately.  As one can see, the changes of $\Delta E$ versus the
number of Fe atoms, $N_{Fe}$, on each of the five sublattices are linear. In particular,
increasing the amount of iron on sublattices A and D leads to a reduction in the $\Delta E$
whereas for other sublattices to its increase.  A similar correlation could be observed
for the cases in which the change was made on two sublattices at the same time without
changing the total concentration of the alloy.  An example of the latter is shown in FIG. 1b.
The results are in line with those presented in FIG. 1a, and they lead to a conclusion that
the configuration $\sigma$-21373 is not the most energetically favorable one
for the $\sigma$-FeCr. Moreover, as it is clear from FIG. 1, a reduction of
$\Delta E$ can be achieved by increasing the population of Fe atoms on the
sublattices A and D or Cr atoms on sublattices B, C and E.
Such changes of the sublattice occupancy lead to the configuration $\sigma$-20080, which was
found as the most favorable energetically in Ref.\onlinecite{Pavlu10}.  Based on the above outlined analysis
we can conclude that if the $\sigma$-phase were created at 0K (the temperature
for which the calculations were carried out), then the sublattice occupancy
would be different from that observed experimentally.

\begin{figure*}[tb]
\includegraphics[width=.99\textwidth]{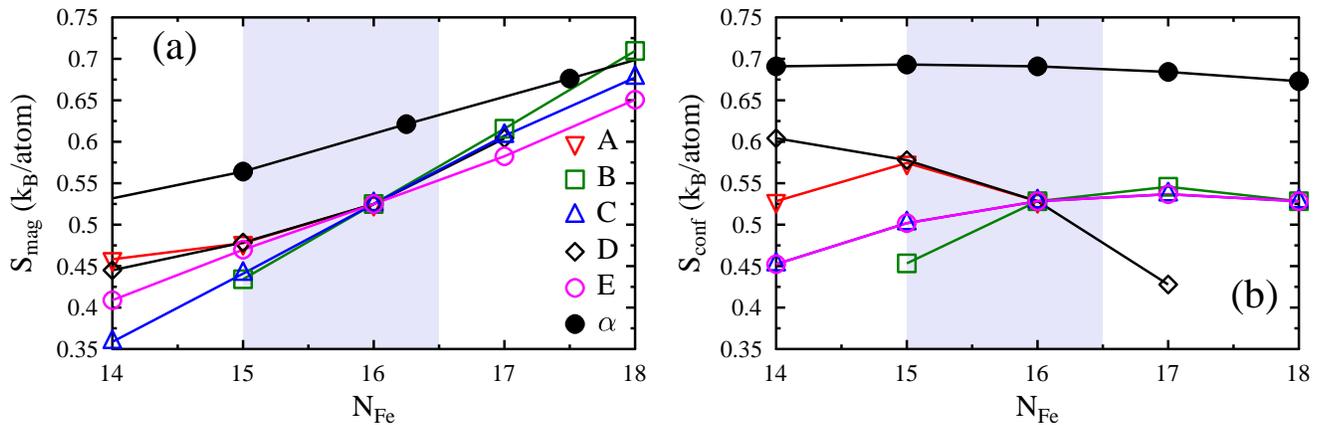}
\caption{(Online color)
Magnetic, $S_{mag}$, (a) and configurational, $S_{conf}$, (b) entropies per one atom for the $\alpha$-phase and for different sublattices of the $\sigma$-phase versus
 the number of Fe-atoms per unit cell, $N_{Fe}$. Solid lines are only to guide the eye.
}
\label{fig2}
\end{figure*}

One must remember that the $\sigma$-phase is a disordered system and each
configuration set presented in FIG. 1 (as well as many others not taken
into account in this calculations) occurs in real terms with various
probabilities.  Therefore, in order to make a fair comparison between our results and the experimental ones
it seems appropriate to calculate the average values of $\Delta E$ as a function of the
concentration of Fe atoms.  The average values, $<\Delta E>$, were obtained based on the
calculated values of $\Delta E$ for each configuration weighted by the probability of its
occurrence.  As can be clearly seen in FIG. 1a, the average $\Delta E$ for the $\sigma$-phase reaches its minimum almost
in the middle of the range where the $\sigma$-phase was found experimentally.

As was mentioned earlier, the $\sigma$-phase in the Fe-Cr system can be formed in the range of
temperatures between $\sim500$ and $\sim830^\circ$C. Annealing above $\sim830^\circ$C leads to
a recovery of the $\alpha$-phase, whereas below $\sim500^\circ$C $\sigma$-phase remains stable.
A comparison between the $\sigma$- and $\alpha$-phase average formation energies is presented in FIG. 1a.
The $<\Delta E>_\alpha$ has been calculated using the KKR-CPA method.  As one can see,
its values are significantly lower than the values found for the $\sigma$-phase in the
corresponding concentration range.  This finding is consistent
with the experimental result that the solid phase of Fe-Cr
obtained from the liquid phase is always of the $\alpha$-type.

In order to obtain a deeper insight into a relationship between the energies of formation of
the $\alpha$- and $\sigma$-phases, one should not restrict itself to a comparison of the
free enthalpy of these phases. Instead, one should also consider the free energy, $F=E-TS$.
To this end, it becomes necessary to estimate the value of the entropy, $S$.

The total entropy which should be taken into account here has several contributions, namely:
configurational entropy, $S_{conf}$,
magnetic entropy, $S_{mag}$,
electronic entropy, $S_{el}$, and
phonon entropy, $S_{ph}$.


The magnetic entropy is one of the quantities that can be determined separately for each of
the discussed atomic configurations. Its contribution to the total entropy was calculated from
the formula
\begin{equation}
  S_{mag} = k_B\sum x_i\ln(2\mu_{i}^{Fe}+1)+ (1-x_i)\ln(2\mu_{i}^{Cr}+1)
\label{eqSM}
\end{equation}
where $k_B$ is Boltzmann constant,
$x_i$ stands for the concentration of Fe on each sublattice,
and the $\mu_i$ is the average magnetic moment of Fe/Cr atoms belonging to this sublattice.
Calculated in this work $\mu$-values were found to be consistent with those already published
\cite{
Soulairol10,   
Lavrentiew10}. 
Consequently, $S_{mag}$-values were
obtained using eq. \ref{eqSM} and are presented in FIG. 2a.
A similar for each site and monotonic behavior of $S_{mag}$ with increasing number of Fe-atoms, $N_{Fe}$,
reflects the fact that the average magnetic moment changes with a
change of Fe-concentration on each sublattice in a similar way.
The magnetic entropy calculated for the $\alpha$-phase was found to be larger then the corresponding one
of the $\sigma$-phase, but the concentration dependence for the two phases is quite similar.
Since the influence of $S_{mag}$ on the phases formation should depend on the difference
of the entropies, $S_{mag_\alpha}-S_{mag_\sigma}$,
one should expect that this type of entropy does not
influense the concentration range of $\alpha-\sigma$ transformation too much, because the difference
$S_{mag_\alpha}-S_{mag_\sigma}$ for various $x$ is only weakly dependent on the concentration.


The configurational contribution to the entropy, $S_{conf}$, was calculated according to the
known formula:
\begin{equation}
  S_{conf} = -k_B\sum x_i\ln(x_i)+ (1-x_i)\ln(1-x_i)
\label{eqSC}
\end{equation}

As in the case of $S_{mag}$, the configurational entropy can
also be determined here for each atomic configuration separately.
Plots presented in FIG. 2b pertinent to the $\sigma$-phase reflect its complex structure. The corresponding values for the $\alpha$-phase
are larger and stay nearly constant in this range of concentration. Based on the results one should expect that these two types of entropy should have rather different influence
on the free energy $F$.


The electronic contribution to the entropy, $S_{el}$, can be determined from the
temperature dependence of the electronic part of the specific heat, $C_{v}$.
Unfortunatelly the frequently used formula viz. $C_v = \beta T+ \gamma T^3+ ...$
that is valid as a low temperature approximation, cannot be used here because the temperature
of $\sim500-800^\circ$ C
at which the $\sigma$-phase can be formed is
almost twice higher than the Debye temperature, $T_D$,
for the $\sigma$-phase (410-480 $^\circ$C)\cite{Cieslak05}
and for the $\alpha$-phase in the discussed range of concentration ($\sim400^\circ$C)\cite{Dubiel10b}
as determined experimentally.
Since the high temperature changes of this type of the specific heat
($\sim \beta T$) are expected to be much weaker than other types
(e.g. phonon contribution, $\sim \gamma T^3$) on one hand, and
because of a saturation character of $C_v(T)$, on the other,
we are not able to take it correctly into account at this stage of calculations.


The phonon contribution, $S_{ph}$, was calculated only for one configuration of atoms
in an ordered unit cell, namely the one corresponding to the $\sigma$-21373 \cite{Dubiel10a}.
Unfortunately, such calculations
are time-consuming and their performation for about 300 atomic arangements, a basis for this
work, was in practice not feasible.  The total $S_{ph}$ was also determined experimentally on the
basis of a phonon spectrum measured on the $\sigma$-phase samples for three different concentrations.
In that case the difference $S_{ph_\sigma}-S_{ph_\alpha} \approx 0.08k_B$, so it is much less than
that for the other types of entropy. Unfortunatelly, the measured values correspond to the average entropy only, and
as such they cannot be used for a careful analysis of the influence of the atomic configurations on the phase transformation.

\begin{figure}[bt]
\includegraphics[width=.49\textwidth]{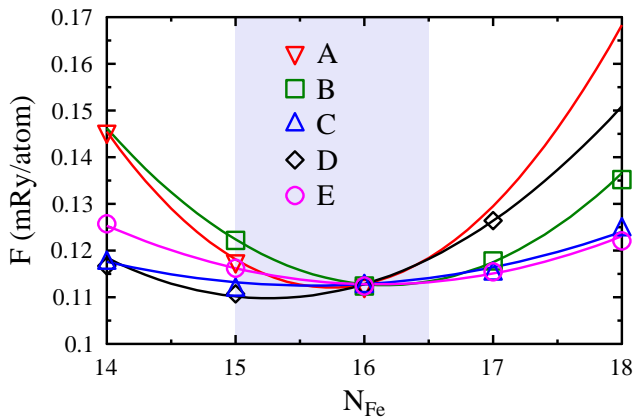}
\caption{(Online color)
The free energy per one atom, $F$, calculated for T = 2000 K as a function of $N_{Fe}$. Only the configurational entropy was taken into account.
}
\label{fig3}
\end{figure}

The above oulined analysis shows that at this stage of computations one has only one type of
entropy, namely the configurational one, that can be quantitatively used to analyse
the behavior of the free energy versus temperature.  Exemplary results obtained in this study
for T = 2000 K are shown in FIG. 3.  It is evident that for each sublattice the free energy
reaches its minimum value in the range of concentration in which the
$\sigma$-FeCr is observed experimentally. Since this phase can be formed only at elevated
temperatures, the sublattice occupancy should corespond to the minimum of $F$ at that
temperatures, as it is indicated by a shadowed-band.

The temperature of 2000 K is by a factor of two larger than the one at which the formation of sigma occurs.
However, the calculations were performed taking into account
the configurational entropy only, which is only one of several contributions to the
total entropy.  Taking into account all types of the entropy will hopefully result in the
lowering of the temperature to the values at which the phase really forms.

In summary, the results presented in this paper can be concluded as follows: (a) if the $\sigma$-phase could be formed at 0 K
the arrangement of atoms would be different (viz. $\sigma$-20080) than the one observed experimentally, (b) the average value of the formation energy
at 0 K has its minimum within the composition where the $\sigma$-phase occurs, (c) the formation energy for the $\alpha$-phase is lower
than the one for the $\sigma$-phase, and consequently always the former precipitates from the {\it liquidus}, and (d) among different possible contributions
to the entropy, the configurational one seems to have the most significant effect on the site occupancy.
\begin{acknowledgments}
This work, supported by the European Communities under the contract of
Association between EURATOM and IPPLM, was carried out within the framework
of the European Fusion Development Agreement. It was also supported by the
Ministry of Science and Higher Education, Warsaw (grant No. N N202 228837)
\end{acknowledgments}


\begin{thebibliography}{58}


\bibitem{Hall66}
\bibinfo{author}{{E.~D. Hall}} and
  \bibinfo{author}{{S.~H. Algie}},
  \bibinfo{journal}{Metall. Rev.} \textbf{\bibinfo{volume}{11}},
  \bibinfo{pages}{61} (\bibinfo{year}{1966}).

\bibitem{Joubert08}
  \bibinfo{author}{{J.-M. Joubert}}
  \bibinfo{journal}{Progr. Mater. Sci.} \textbf{\bibinfo{volume}{53}},
  \bibinfo{pages}{528} (\bibinfo{year}{2008}).

\bibitem{Cieslak08c}
  \bibinfo{author}{{J. Cieslak}},
  \bibinfo{author}{{M. Reissner}},
  \bibinfo{author}{{S.~M. Dubiel}},
  \bibinfo{author}{{J. Wernisch}}
  and
  \bibinfo{author}{{W. Steiner}},
  \bibinfo{journal}{J. Alloys Comp.} \textbf{\bibinfo{volume}{460}},
  \bibinfo{pages}{20} (\bibinfo{year}{2008}).

\bibitem{Turchi94}  
        \bibinfo{author} {{P. E. A. Turchi    }},
        \bibinfo{author} {{L.   Reinhard  }},
             and
        \bibinfo{author} {{G. M. Stocks    }},
        \bibinfo{journal}{Phys. Rev. B}
\textbf{\bibinfo{volume} {50}},
        \bibinfo{pages}  {15542}
       (\bibinfo{year}   {1994}).

\bibitem{Olsson03} 
        \bibinfo{author} {{P.   Olsson    }},
        \bibinfo{author} {{I. A.Abrikossov}},
        \bibinfo{author} {{L.   Vitos     }},
             and
        \bibinfo{author} {{J.   Wallenius }},
        \bibinfo{journal}{J. Nucl. Mater},
\textbf{\bibinfo{volume} {321}},
        \bibinfo{pages}  {84}
       (\bibinfo{year}   {2003}).

\bibitem{Korzhavyi05}  
        \bibinfo{author} {{P. A. Korzhavyi }},
        \bibinfo{author} {{Bo    Sundman   }},
        \bibinfo{author} {{M.    Selleby   }},
             and
        \bibinfo{author} {{B.    Johansson }},
        \bibinfo{journal}{Mater. Res. Soc. Symp. Proc.}
\textbf{\bibinfo{volume} {842}},
        \bibinfo{pages}  {S4.10}
       (\bibinfo{year}   {2005}).

\bibitem{Ackland09} 
        \bibinfo{author} {{G. J.Ackland   }},
        \bibinfo{journal}{Phys. Rev. B}
\textbf{\bibinfo{volume} {79}},
        \bibinfo{pages}  {094202}
       (\bibinfo{year}   {2009}).

\bibitem{Kabliman09}
  \bibinfo{author}{{E.~A. Kabliman}},
  \bibinfo{author}{{A.~A. Mirzoev}}
  and
  \bibinfo{author}{{A.~L. Udovskii}},
  \bibinfo{journal}{Phys. Met. Metallogrephy} \textbf{\bibinfo{volume}{108}},
  \bibinfo{pages}{435} (\bibinfo{year}{2009}).

\bibitem{Cieslak08b}
  \bibinfo{author}{{J. Cieslak}},
  \bibinfo{author}{{J. Tobola}},
  \bibinfo{author}{{S.~M. Dubiel}},
  \bibinfo{author}{{S. Kaprzyk}},
  \bibinfo{author}{{W. Steiner}}
  and
  \bibinfo{author}{{M. Reissner}},
  \bibinfo{journal}{J. Phys.: Condens. Matter.} \textbf{\bibinfo{volume}{20}},
  \bibinfo{pages}{235234} (\bibinfo{year}{2008}).

\bibitem{Cieslak10a}
  \bibinfo{author}{{J. Cieslak}},
  \bibinfo{author}{{J. Tobola}},
  and
  \bibinfo{author}{{S.~M. Dubiel}},
  \bibinfo{journal}{Phys.\ Rev.\ B} \textbf{\bibinfo{volume}{81}},
  \bibinfo{pages}{174203} (\bibinfo{year}{2010}).

\bibitem{Sluiter75}
  \bibinfo{author}{{M.H.F. Sluiter}},
  \bibinfo{author}{{K. Esfarjani}}
  and
  \bibinfo{author}{{Y. Kawazoe}},
  \bibinfo{journal}{Phys.\ Rev.\ Lett.} \textbf{\bibinfo{volume}{75}},
  \bibinfo{pages}{3142} (\bibinfo{year}{1995}).

\bibitem{Pavlu10}
  \bibinfo{author}{{J. Pavl\.u}},
  \bibinfo{author}{{J. V\v{r}e\v{s}\v{t}\'{a}l}}
  and
  \bibinfo{author}{{M. \v{S}ob}},
  \bibinfo{journal}{Intermetallics} \textbf{\bibinfo{volume}{18}},
  \bibinfo{pages}{212} (\bibinfo{year}{2010}).

\bibitem{Soulairol10}  
        \bibinfo{author} {{R.   Sourairol }},
        \bibinfo{author} {{Chu-Chun} {Fu          }},
             and
        \bibinfo{author} {{C.   Barretteau}},
        \bibinfo{journal}{J. Phys. Cond. Matter}
\textbf{\bibinfo{volume} {22}},
        \bibinfo{pages}  {295502}
       (\bibinfo{year}   {2010}).

\bibitem{Lavrentiew10} 
        \bibinfo{author} {{M. Yu. Lavrentiev}},
        \bibinfo{author} {{D.   Nguyen-Manh}},
             and
        \bibinfo{author} {{S. L. Dudarev}},
        \bibinfo{journal}{Phys. Rev. B}
\textbf{\bibinfo{volume} {81}},
        \bibinfo{pages}  {184202}
       (\bibinfo{year}   {2010}).

\bibitem{Cieslak05}
  \bibinfo{author}{{J. Cieslak}},
  \bibinfo{author}{{B.F.O. Costa}},
  \bibinfo{author}{{S.~M. Dubiel}},
  \bibinfo{author}{{M. Reissner}}
  and
  \bibinfo{author}{{W. Steiner}},
  \bibinfo{journal}{J. Phys.: Condens. Matter.} \textbf{\bibinfo{volume}{17}},
  \bibinfo{pages}{6889} (\bibinfo{year}{2005}).

\bibitem{Dubiel10b}
  \bibinfo{author}{{S.~M. Dubiel}},
  \bibinfo{author}{{J. Cieslak}}
  and
  \bibinfo{author}{{B.F.O. Costa}},
  \bibinfo{journal}{J. Phys.: Condens. Matter.} \textbf{\bibinfo{volume}{22}},
  \bibinfo{pages}{055402} (\bibinfo{year}{2010}).

\bibitem{Dubiel10a}
  \bibinfo{author}{{S.~M. Dubiel}},
  \bibinfo{author}{{J. Cieslak}},
  \bibinfo{author}{{W. Sturhahn}},
  \bibinfo{author}{{M. Sternik}},
  \bibinfo{author}{{P. Piekarz}},
  \bibinfo{author}{{S. Stankov}}
  and
  \bibinfo{author}{{W. Parlinski}},
  \bibinfo{journal}{Phys. Rev. Lett.} \textbf{\bibinfo{volume}{104}},
  \bibinfo{pages}{155503} (\bibinfo{year}{2010}).


\end{thebibliography}
\end{document}